\documentclass[a4paper]{raa}           
\usepackage{graphicx,times}
\usepackage{natbib}
\usepackage{amssymb,amsmath}
\bibpunct{(}{)}{;}{a}{}{,}

\usepackage{hyperref}

\hypersetup{colorlinks = true, linkcolor = green, anchorcolor = red, citecolor = blue, filecolor = red, pagecolor = red, urlcolor = red}

\begin{document}

   \title{Near-infrared studies of nova V5584 Sgr in the pre-maximum and early decline phase
$^*$
% \footnotetext{\small $*$ Supported by the National Natural Science Foundation of China.}
}

 \volnopage{ {\bf 2014} Vol.\ {\bf X} No. {\bf XX}, 000--000}
   \setcounter{page}{1}

   \author{Ashish Raj\inst{1}, D. P. K. Banerjee\inst{2}, N. M. Ashok\inst{2}, Sang Chul KIM
      \inst{1,3} }
%% Here is an example of three authors come from different institutes.
%% For single author or all the authors from an institute, use "\inst{}" only

   \institute{Korea Astronomy and Space Science Institute (KASI), Daejeon 305-348, Korea; {\it ashish@kasi.re.kr}\\
%% Please give the E-mail address of the author, to whom future correspondence and
%% offprint requests will be sent.
        \and
             Physical Research Laboratory, Navarangpura, Ahmedabad-380009, India.\\
             \and Korea University of Science and Technology (UST), Korea.\\
% 	\and
% %	  Center for Astrophysics, University of Science and Technology of China, Hefei 230026, China\\
% Key Laboratory for Research in Galaxies and Cosmology, The University of Science
% and Technology of China, Chinese Academy of Sciences, Hefei, Anhui, 230026, China\\
% \and 
% Polar Research Institute of China,
% Jinqiao Rd. 451, Shanghai, 200136, China\\
\vs \no
   {\small Received  ; accepted }
}

\abstract {We present near-infrared spectroscopic and photometric observations of nova V5584 Sgr taken during the first 12 days following its discovery on 
Oct. 26.439 UT 2009. The evolution of the spectra is shown from the initial P Cygni phase to an emission line phase. The prominent carbon lines seen in the 
$JHK$ spectra closely match those observed in a FeII class nova outburst. The spectra show first-overtone CO bands in emission between 2.29-2.40 $\mu$m. By examining WISE and other publicly available data, we show that the nova underwent a pronounced dust formation phase during February - April 2010.
}
 
 \authorrunning{Raj et al. }            %author_head in even pages
   \titlerunning{Near-infrared studies of nova V5584 Sgr}  % title_head in odd pages
   \maketitle

%________________________________________________ sections below
%
\section{Introduction}           %% first-level sections will be auto-capitalized
% \label{sect:intro}

Nova Sagittarii 2009 No. 4 (V5584 Sgr) was discovered on Oct. 26.439 UT 2009 by Nishiyama and Kabashima on two 60s unfiltered 
CCD frames at a magnitude of 9.3 in V (\cite{Nishiyam09}). Nothing was visible at this position on their two survey frames taken respectively on Oct 20.449 UT 2009 with limiting
magnitude 13.9 and
Oct. 21.451 UT 2009 with limiting magnitude 13.4. \cite{Corelli09} has reported that nothing was visible at the nova position on the Palomar plate 
(limiting magnitude 21) taken on Oct. 26.764 UT 2008. 

A low-resolution optical spectrum obtained by \cite{Kinugasa09} on Oct. 27.4 UT 2009, with the 1.5m telescope at Gunma Astronomical Observatory  showed 
hydrogen Balmer series absorption lines, with H$\alpha$ having a prominent P-Cygni profile and a full width at half-maximum (FWHM) of about 600 km s$^{-1}$ 
suggesting the object to be a nova in its early stage.
The absorption minimum of the H$\alpha$ was blue-shifted by 900 km s$^{-1}$ from the emission peak.
Another low-resolution spectrum obtained by \cite{Maehara09} on Oct. 27.42 UT 2009 also showed a similar profile of
the H$\alpha$ line, suggesting that the object is a classical nova. 
\cite{Munari09} have taken low-, medium-, and high-resolution spectra for this nova with 
the 0.6m telescope of the Schiaparelli Observatory in Varese. 
The low- and medium-resolution spectra taken on Oct. 28.73 UT 2009 show a well developed and highly reddened absorption continuum. The FWHM of the
absorption lines are $\sim$ 310 km s$^{-1}$. 
The heliocentric radial velocity of the absorption lines were $\sim$ -283 km s$^{-1}$ and the separation between the 
absorption and emission components was 440 km s$^{-1}$. The high resolution echelle spectra taken on Oct. 29.72 UT 2009 show very weak emission from Balmer 
and Fe II multiplets. They report V magnitudes of 9.74, 9.31 and 9.21 on Oct. 27.719 UT, 28.709 UT 
 and 29.708 UT 2009, 
respectively. \cite{Munari09} have pointed out that the object is a nova of Fe II class and approaching the optical maximum. 

The near-infrared (NIR) spectrum taken by \cite{Raj09} on Oct. 29.58 UT 2009 showed the strong P-Cygni profiles for H I, O I, C I 
and N I lines. 
Subsequent observations show strengthening of the emission components and the first overtone CO bands was detected in the spectrum taken on
Nov. 5.64 UT 2009. 
The near-IR observations obtained on Feb. 10 UT 2010 by \cite{Russell10}, after V5584 Sgr came out of the solar conjunction, showed the dust 
formation in the nova ejecta and the dust temperature was estimated as 880 $\pm$ 50 K.
The optical spectra taken on Jun. 4 and Aug. 10 UT 2010 by \cite{Poggiani11} show that the nova had entered the nebular phase.

\section{Observations}

The near-IR observations were obtained using the 1.2m telescope of Mt. Abu Infrared Observatory from Oct. 28.59 to Nov. 8.56 UT 2009 covering one epoch before 
the optical maximum and the early decline phase. The spectra were obtained at a resolution of $\sim$ 1000 using a Near-Infrared Imager/Spectrometer with a 
256 $\times$ 256 HgCdTe NICMOS3 array. Spectral calibration was done using the OH sky lines that register with the stellar spectra. The spectra of the comparison 
star SAO 161520 (spectral type A1V; effective temperature 9230 K) were taken at similar airmass as that of V5584 Sgr to ensure that the ratioing process 
(nova spectrum divided by the standard star spectrum) removes the telluric features reliably. The H I absorption lines in the spectra of standard star were 
removed manually  before ratioing to avoid artificially generated emission lines in the ratioed spectrum. The ratioed spectra were then multiplied by a blackbody 
curve corresponding to the standard star's effective temperature to yield the final spectra.

\begin{table*}
\scriptsize
%\tablewidth{0pt}
\small
% \bc
%\begin{minipage}[]{100mm}

%\centering
\caption{A log of near-infrared spectroscopic and photometric observations of V5584 Sgr. The optical maximum is assumed to be its detection date. 
The date of optical maximum is taken as Oct. 29.71 UT 2009.}
 \small
\begin{tabular}{ccccccccccc}
\hline
Date of      &Days since    &\multicolumn{3}{|c|}{Integration time (s)} &\multicolumn{3}{|c|}{Integration time (s)} &\multicolumn{3}{|c|}{Nova Magnitude} \\
Observation (UT)  &optical max             &J-band      &H-band   &K-band  &J-band      &H-band   &K-band   &J-band      &H-band   &K-band  \\
\hline
   &   &\multicolumn{3}{|c|}{Spectroscopic Observations}   &\multicolumn{4}{|l|}{Photometric Observations}  \\
\hline
2009 Oct. 28.59    &-1.12&--  &--   &--   &250 &220  &105 &6.97$\pm$0.04 &6.65$\pm$0.04  &6.45$\pm$0.07 \\
2009 Oct. 29.58    &-0.13&60  &50   &90  &25  &110  &50 &-- &-- &6.28$\pm$0.06 \\
2009 Oct. 30.58    &0.87   &--  &--   &--  &75  &110 &30 &6.92$\pm$0.05 &6.53$\pm$0.03 &6.33$\pm$0.04\\
2009 Nov. 02.59  &3.88    &--  &--   &--  &50  &110  &105  &6.90$\pm$0.02&6.64$\pm$0.04  &6.37$\pm$0.06\\
2009 Nov. 03.59  &4.88    &120  &120   &--  &--  &--  &--  &--&--  &--\\
2009 Nov. 04.60    &5.89&90 &--   &--  &75 &110 &30 &7.20$\pm$0.06 &6.88$\pm$0.05  &6.45$\pm$0.04 \\
2009 Nov. 05.64    &6.93&90 &90   &90  &-- &-- &-- &-- &--  &-- \\
2009 Nov. 06.58    &7.87   &90  &90   &--   &75  &110  &50 &7.36$\pm$0.03 &7.16$\pm$0.02  &6.83$\pm$0.06 \\
2009 Nov. 07.61    &8.90 &--  &--   &--   &100 &50 &-- &7.84$\pm$0.02 &7.60$\pm$0.08  &--\\
2009 Nov. 08.56  &9.85  &120 &120  &120  &125  &75  &55  &7.86$\pm$0.04 &7.66$\pm$0.03  &7.38$\pm$0.07\\
% 2010 Apr. 11.94    &164.23 &-- &--  &-- &1000  &1100  &105  &12.81$\pm$0.08 &10.33$\pm$0.06  &10.10$\pm$0.16\\
% 2010 Apr. 23.91    &176.20  &-- &--  &-- &2000 &270  &55  &13.48$\pm$0.11 &10.40$\pm$0.01  &9.57$\pm$0.14\\
\hline
\end{tabular}
\label{table1}
%\end{minipage}
\end{table*}

The photometry was done in clear sky conditions using the NICMOS3 array in the imaging mode in $JHK$ bands. Several frames were obtained in all the bands, in 
4 dithered positions, offset by $\sim$ 30 arcsec. The sky frames, which are subtracted from the nova frames, were generated by median combining the dithered 
frames. The star SAO 161520 (with the $JHK$-band magnitudes of 4.803, 4.622 and 4.360, respectively) located close to the nova was used for photometric 
calibration. All data reduction and analysis were done using $IRAF$. The log of the spectroscopic and photometric observations and the $JHK$ magnitudes are 
given in Table 1.

\begin{table}
\caption[]{A list of the lines identified from the $JHK$ spectra. The
  additional lines contributing to the identified lines are listed.}
\begin{tabular}{llll}
\hline\\
Wavelength & Species  & Other contributing  \\
(${\mu}$m) & &lines and remarks   \\
\hline \\

% 1.0830   & He\,{\sc i}         &      \\
1.0938   & Pa $\gamma$        &      \\
1.1287          & O\,{\sc i} &               \\
1.1330          & C\,{\sc i} &               \\
1.1381-1.1404   & Na\,{\sc i}        &    C\,{\sc i} 1.1415  \\
1.1600-1.1674   & C\,{\sc i}  & strongest lines at 1.1653,\\
                &             &           1.1659,1.16696    \\
% 1.6872          & Fe\,{\sc ii}&                             \\
1.1748-1.1800   & C\,{\sc i}  & strongest lines at 1.1748,  \\
                &             &          1.1753,1.1755      \\
1.11819-1.1896   & C\,{\sc i}  &   \\ 
1.1819-1.2614   & C\,{\sc i}, N\,{\sc i}  &  blend of sveral C\,{\sc i} and N\,{\sc i} lines \\
% 1.1828          & Mg\,{\sc i} &               \\
% 1.1880          & C\,{\sc i} &               \\
% 1.2074          & N\,{\sc i} &               \\
% 1.2187,1.2204          & N\,{\sc i} &               \\
% 1.2249,1.2264          & C\,{\sc i} &               \\
%1.2329          & N\,{\sc i} &               \\
% 1.2382          & N\,{\sc i} &               \\
1.2461,1.2469 & N\,{\sc i} & blended with O\,{\sc i} 1.2464 \\
1.2562,1.2569 & C\,{\sc i} & blended with O\,{\sc i} 1. 2570 \\
1.2818   & Pa $\beta$         &     \\
% 1.2950   & C\,{\sc i}         &     \\
1.3164   & O\,{\sc i}         &     \\
% 1.5040   & Mg\,{\sc i}        &  blended with Mg\,{\sc i} 1.5025,\\
1.5256   & Br 19              &           \\
1.5341   & Br 18              &           \\
1.5439   & Br 17              &           \\
1.5557   & Br 16              &           \\
1.5701   & Br 15              &           \\
1.5749   & Mg\,{\sc i}        & blended with Mg\,{\sc i} 1.5741,  \\
         &                    &           1.5766,C\,{\sc i} 1.5788 \\
1.5881   & Br 14              &  blended with C\,{\sc i} 1.5853    \\
1.6005   & C\,{\sc i}         &           \\
1.6109   & Br 13              &           \\
1.6407   & Br 12              &           \\
1.6806   & Br 11              &           \\
1.6890   & C\,{\sc i}         &           \\
% 1.7002   & He\,{\sc i}         &           \\
1.7045   & C\,{\sc i}         &           \\
1.7109   & Mg\,{\sc i}        &               \\
% 1.7234-1.7275 & C\,{\sc i}    & several C\,{\sc i} lines  \\
1.7362   & Br 10              &  affected by C\,{\sc i} 1.7339 line    \\
1.7449 & C\,{\sc i}           &           \\
1.7605-1.7638 & C\,{\sc i}    &           \\
% 2.0581 & He\,{\sc i}          &           \\
% 2.1120,2.1132   & He\,{\sc i}         &           \\
2.1156-2.1295 & C\,{\sc i}    &           \\
% 2.1452 & Na\,{\sc i}          &           \\
2.1655   & Br $\gamma$        &           \\
2.2056,2.2084 & Na\,{\sc i}   &           \\
2.2156-2.2167 & C\,{\sc i}    &           \\
2.29-2.40           & CO               & $\Delta$v=2 bands             \\
2.2906   		    & C \,{\sc i}      	& 				\\
2.3130              & C \,{\sc i}       &               \\
2.3348              & Na \,{\sc i}       &               \\
2.3379            & Na \,{\sc i}       &               \\
\hline
\end{tabular}
\label{table4}
\end{table}
   
\section{Results}

We describe the key results in the following sub-sections.

\subsection{General characteristics of $V$ and $JHK$ light curves}

The $V$ band light curve based on the data from AAVSO and the $JHK$ band light curves from Mt. Abu Observatory are 
presented in Figures 1 and 2, respectively. The nova reached 
a maximum value of $V_{max}$ = 9.2 on Oct. 29.71 UT 2009.
We estimate $t_2$ (time taken by the nova to decline by 2 magnitudes from the optical maximum) to be 25 $\pm$ 
1 d from a least square regression fit to the post maximum light curve. We determine the absolute magnitude of the nova to be 
M$_V$ = -7.7 $\pm$ 0.2 using the maximum magnitude versus rate of decline (MMRD) relation of \cite{Della95}. We derive the reddeing  $E(B - V)$ = 0.94 
from \cite{Schlafly11} towards the direction of the nova which gives  interstellar extinction A$_V$ = 2.9 for R = 3.1. Using the distance modulus relation, we obtain a value of the distance $d$ = 6.3 $\pm$ 0.5 kpc to the nova. 
Using this value of $d$ and the Galactic latitude of the nova -3.1 deg, we estimate the height of the nova to be $z$ = 341 $\pm$ 30 pc below the Galactic plane. 
The outburst bolometric luminosity of V5584 Sgr as calculated from M$_V$ is L$_O$ $\sim$ 1.1 $\times$ 10$^5$ L$_\odot$. 
As there are no observations after $\sim$ 30 days from the discovery date due to the solar conjuction of the 
nova 
we estimate the 
time taken by the nova to decline by 3 magnitudes from the optical maximum $t_3$ = 46 $\pm$ 1 d by 
using the relation $t_3$ = 2.7($t_{2}$)$^{0.88}$ of \cite{Warner95}. The observed value of the outburst amplitude (difference between the limiting magnitude of 
detection and the
optical maximum magnitude) $\bigtriangleup V$ = 12 and $t_{2}$ = 25 days for V5584 Sgr
is consistent with the outburst amplitude versus decline rate plot for classical novae presented by \cite{Warner08} which shows $\bigtriangleup V$ = 10 - 13 for 
$t_{2}$ = 25
days.

\begin{figure}
\begin{center}
\includegraphics[width=3.3in,height=2.5in,clip]{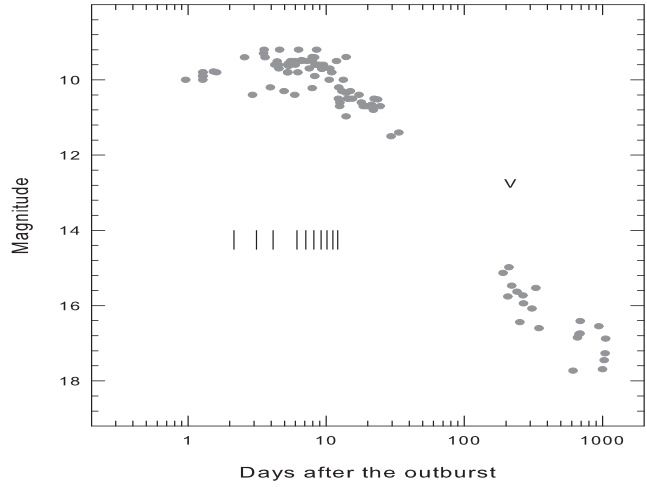}
\caption[The $V$ and $JHK$ band light curves of KT Eri]{The $V$ band light curve of V5584 Sgr from AAVSO data. 
The days of
spectroscopic and photometric observations are shown as solid lines below.
}
\label{ch5_2}
\end{center}
\end{figure}

A classification scheme of novae based on the optical light curves is presented by \cite{Strope10}. 
Such a classification is based on the post-maximum time when the light curve declined by 3.0 magnitudes, namely $t_{3}$ 
and the shape of the light curve. We classify the optical light curve of 
V5584 Sgr as D(46) according to the estimated value of $t_{3}$ (46 days), where D(46) denotes the dust formation as discussed later in subsection 3.5.

\begin{figure}
\begin{center}
\includegraphics[width=3.3in,height=2.5in,clip]{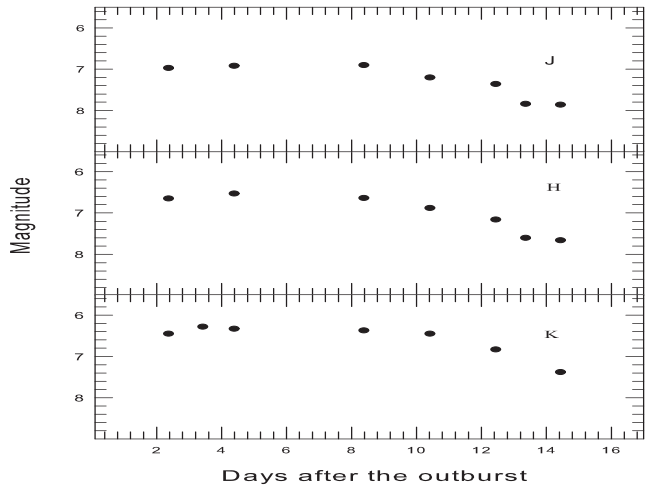}
\caption[The $V$ and $JHK$ band light curves of KT Eri]{The Mt.Abu $JHK$ band light curves of V5584 Sgr.}
\label{ch5_2}
\end{center}
\end{figure}

\subsection{Line identification, evolution and general characteristics of the $JHK$ spectra}

The $JHK$ spectra are presented in Figures 3 to 5 respectively and the line identification in graphical and tabular form are given in Fig. 6 and Table 2, respectively.
The near-IR observations presented here cover the pre-maximum light to early post-maximum decline. The first infrared spectra  taken on Oct. 29.58 UT 2009 are 
dominated by lines of hydrogen, neutral nitrogen, carbon and oxygen with prominent P-Cygni profiles. The FWHM of the absorption and emission components
of the Pa$\beta$ line are 400 and 560 km s$^{-1}$ and the separation between the absorption and emission peaks for all the lines is typically in the range of 550-650 km s$^{-1}$.
The next spectra taken on Nov. 3.59 UT 2009 show considerable strengthening of the emission components having P-Cygni absorption with reduced intensity. Spectral templates in the NIR for the characteristic spectra of the
Fe II and He/N  class of novae have been presented  by \cite{Banerjee12}. The essential NIR spectral features 
that distinguish between the Fe II and He/N class of novae are the strong carbon lines in the former class. 
The  $JHK$  spectra presented here show the presence of carbon lines at 1.166 $\mu$m and 1.175 $\mu$m in the $ J$ band, 1.689 $\mu$m and several lines between 1.72  $\mu$m and 1.79 $\mu$m in the $H$  band and lines between 2.11 $\mu$m to 2.13  $\mu$m and 2.29  
$\mu$m to 2.31 $\mu$m lines in the $K$ band. Thus the carbon rich spectra  indicate that  V5584 Sgr belongs to Fe II class. The presence of Na and Mg lines 
(e.g., Na I 2.2056$\mu$m, 2.2084$\mu$m) in the spectrum can be regarded as an indicator of dust formation in the nova ejecta (\cite{Das08}) and this
gets further support from the analysis of the 3- to 14-$\mu$m spectroscopy
(\cite{Russell10}).

\subsection{First overtone CO  detection and modeling}

The first K band spectrum taken on Oct. 29.58 UT 2009, very close to optical maximum, does not show evidence for first overtone CO emission at 2.29 $\mu$m and beyond.
It is possible that CO was present at this stage but below or just at the detection level (see \cite{Raj11}). Nevertheless, we notice that the K-band spectrum
taken on Nov. 5.64 UT 2009 shows
clear evidence for first overtone CO emission. This is one of the most interesting result in V5584 Sgr since CO detections are very rare in novae. 
Among more than 300 Galactic novae, only 9 are found to show CO emissions (i.e., $<$ 3$\%$).
The next K band spectra taken on Nov. 8.56 UT 2009 does not cover the CO emission region so it is not possible to
comment on the duration of the CO emission. Based on the detailed studies by \cite{Das09} and \cite{Raj12}, we have compiled in Table 3 the complete list of
novae known to have shown first overtone CO emission. 
Theoretically, \cite{Pontefract04} pointed out that CO should form early after outburst and remain approximately constant in strength for 12-15 days thereafter 
and then get rapidly destroyed.
The early appearance of CO in V5584 Sgr is consistent with the predictions.
It may be mentioned that models for CO emission as developed by \cite{Rawlings88} and 
\cite{Pontefract04} found that the outer parts of the ejecta have to be much denser and less ionized than the bulk of the wind which favors the formation of substantial molecules. Carbon has to be neutral and in such a neutral carbon region, 
the carbon ionization continuum, which extends to less than 1102\AA, shields several molecular species
against the dissociative UV flux from the central star. The relatively denser and cooler environment that is conducive for molecular growth also favors dust 
formation and every CO forming nova has always been known to form dust. However, the converse case has not be seen. The reason for the non detection of CO in other 
dust forming novae is not known. A possible reason for the  non detection of CO emission in novae with dust formation is the paucity of near-infrared spectral 
observations in the early phase of nova evolution and its shorter duration. Alternatively, in these novae either CO did not form for reasons which are not 
clearly understood or it was present but below the detection limit of observations.

\begin{figure}
 \begin{center}
% \centering
 \includegraphics [width=3.0in,height=5.0in,clip]{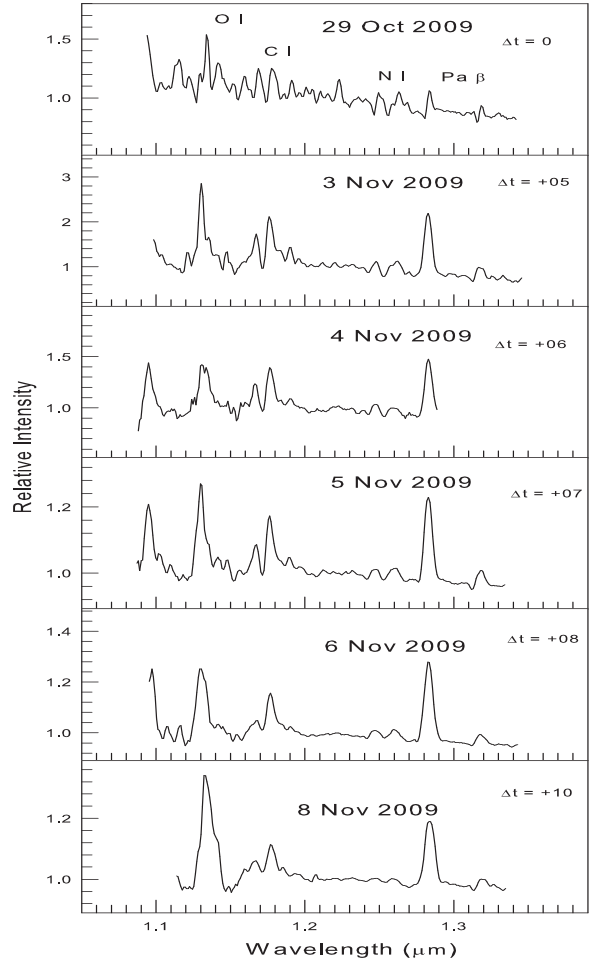}
  \caption[The $J$ band spectra of KT Eri]{The $J$ band spectra of V5584 Sgr are shown at different epochs. The relative intensity is normalized to
unity at 1.25 ${\rm{\mu}}$m. The epoch relative to the optical maximum are given in days for each spectrum.}
 \label{ch5_3}
 \end{center}
 \end{figure}

% ----------- end Figure 2 ----------------------
\begin{figure}
\begin{center}
 \includegraphics[width=3.0in,height=5.0in, clip]{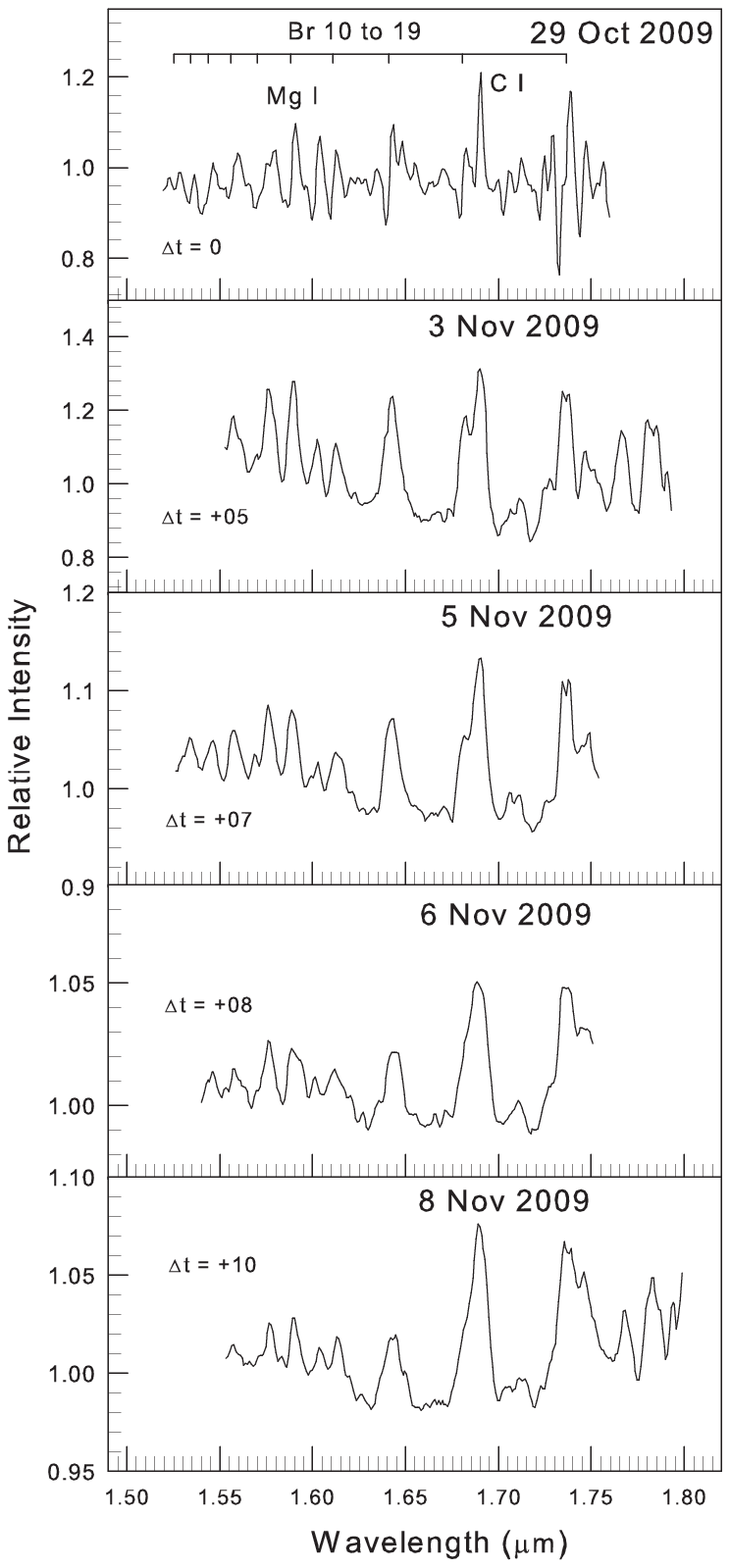}
  \caption[The $H$ band spectra of KT Eri]{The $H$ band spectra of v5584 Sgr are shown at different epochs. The
   relative intensity is normalized to unity at 1.65 ${\rm{\mu}}$m. The epoch relative to the optical maximum are given in days for each spectrum.}
  \label{ch5_4}
  \end{center}
  \end{figure}
%
%% ----------- end Figure 3 ----------------------

%% ----------- begin Figure 4 ----------------------
  \begin{figure}
  \begin{center}
  \includegraphics[width=3.0in,height=5.0in, clip]{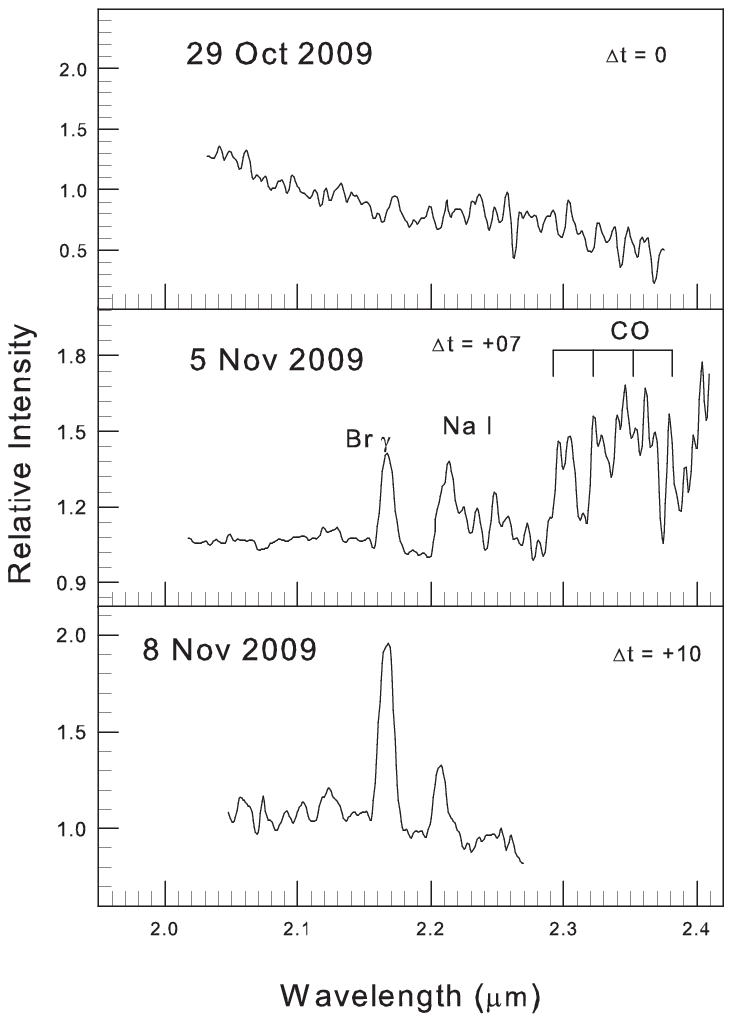}
 \caption[The $K$ band spectra of KT Eri]{The $K$ band spectra of V5584 Sgr are shown at different epochs. The
   relative intensity is normalized to unity at 2.2 ${\rm{\mu}}$m. The epoch relative to the optical maximum are given in days for each spectrum.}
     \label{ch5_5}
     \end{center}
  \end{figure}

\begin{figure}
  \begin{center}
  \includegraphics[width=3.0in,height=5.0in, clip]{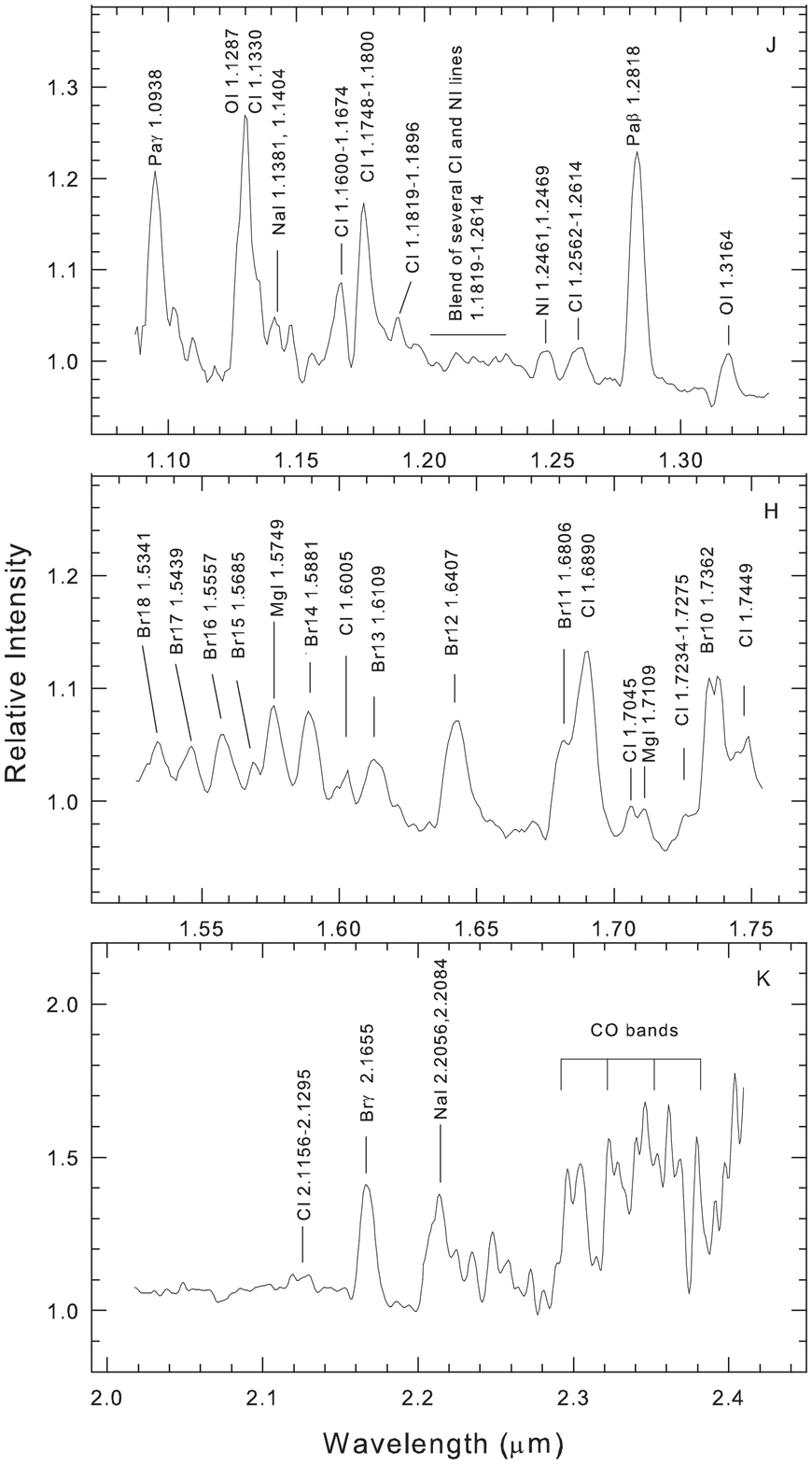}
 \caption[The line identification is shown for the spectra of 2009 November 5 in the $JHK$
bands as given in Table 2.]{The line identification is shown for the spectra of Nov. 5.64 UT 2009 in the $JHK$
bands as given in Table 2.}
     \label{ch5_5}
     \end{center}
  \end{figure}
  
We used the  model developed by \cite{Das09} in the case of V2615 Oph to characterize the CO emission in V5584 Sgr (Fig. 7). Compared to V2615 Oph, the CO 
detection here is only for one epoch and the S/N of the spectrum is poor (approximately 10) and thus not conducive for accurate modeling. However, since CO 
detections are rare, it is desirable to have rough model estimates of the CO parameters even if such estimates are not quite accurate. In the model 
calculations  the CO gas is considered to be in thermal equilibrium with the same temperature for calculating the level populations of rotation and vibration 
bands (see \cite{Das09} for more details). The data  covers only  three of the bands ($\nu$ = 2-0, 3-1, 4-2) and the C I lines at 2.2906 and 2.3130 $\mu$m 
are likely blended with Na I lines at 2.3348 and 2.3379 $\mu$m, thereby giving rise to further complications. 
Allowing for all these factors, we estimate the 
temperature to be 3500 $\pm$ 750 K and constrain the upper limit for the mass of the 
CO gas to be in the range 1-6 $\times$ 10$^{-8}$ M$_\odot$. The errors are fairly large but the central values of the mass and temperature are very similar 
to the values obtained in V2615 Oph (\cite{Das09}) and V496 Sct (\cite{Raj12}). We do not make any attempt to determine the $^{12}$C/$^{13}$C ratio 
given the low S/N ratio of the spectrum.

\begin{figure}
\begin{center}
\includegraphics[width=2.5in,height=2.5in,clip]{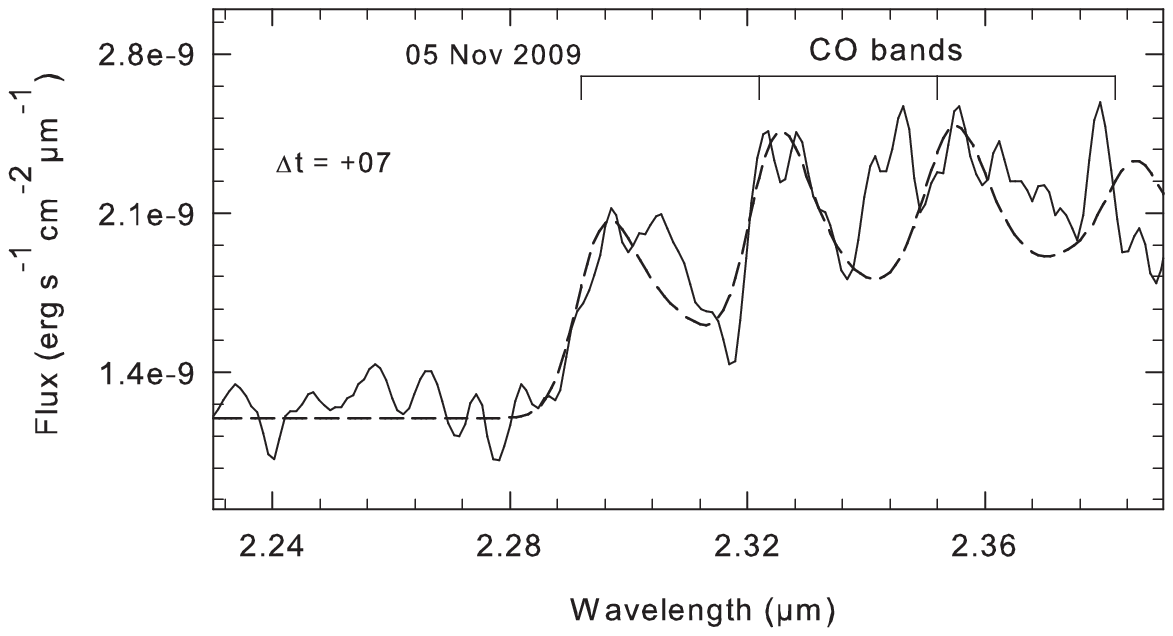}
\caption[] {The model curve is shown as dashed line in comparison with the first overtone CO bands observed in V5584 Sgr. The fit is made
 for a constant CO mass of 3 $\times$ 10$^{-8}$ M$_\odot$ and the temperature of the gas T$_{CO}$ is 3500 K. The epoch relative to the optical maximum are given 
 in days for each spectrum.}
\label{ch5_2}
\end{center}
\end{figure}

\subsection{Fireball phase}

The continuum from a nova's ejecta near maximum light is known to mimic the photosphere of an A-F spectral type star (\cite{Gehrz08}). The spectral energy 
distribution (SED) of the pseudo-photosphere during this fireball stage is generally well approximated by a blackbody. As V5584 Sgr showed pre-maximum rise and 
has a well-defined optical maximum we have constructed the SED to study the fireball phase for the nova. Using the AAVSO data for the 
following optical magnitudes $B$=10.4, $V$=9.5, $R_C$=8.7 and$I_C$=7.6 for Nov. 2 UT 2009 along with the present $JHK$ magnitudes of 
Nov. 2.59 UT 2009, 
we derived the SED in the fireball phase. 
The observed magnitudes were corrected for extinction using \cite{Schlafly11}. We obtain a temperature of $T_{bb}$ = 9000 $\pm$ 500 K from a balckbody fit to the 
SED shown in the top panel of Fig. 8. This is consistent with the A-F spectral type for the pseudo-photospheres displayed by novae at outburst (\cite{Gehrz08}).
Using the relation given by \cite{Ney78}, the blackbody angular diameter $\theta_{bb}$ in arcseconds is calculated, viz, 

\begin{equation}
\nonumber
\theta_{bb} = 2.0 \times 10{^{\rm 11}} (\lambda F_{\lambda})_{max}^{1/2}\times T_{bb}^{-2}
\end{equation}
where $ (\lambda F_{\lambda})_{max}$ = 6.56 $\times$ 10${^{\rm -15}}$ W cm$^{-2}$ and $T_{bb}$ = 9000 K. We obtain a value of 0.2 milliarcsec for the angular diameter.
This value for the angular diameter can be used to estimate the distance to the nova by assuming a constant expansion rate for the ejecta and the relation given
by \cite{Gehrz08}, which follows as

\begin{equation}\nonumber
 d = 1.15 \times 10{^{\rm -3}} (V_{ej})t/ \theta_{bb} %\nonumber
\end{equation}

where d is in kpc, $V_{ej}$ in km s$^{-1}$, t is time after the outburst in days and $\theta_{bb}$ in milliarcsec.
The estimated value of $\theta_{bb}$ will always be a lower limit since it is applicable for a blackbody (\cite{Ney78}; \cite{Gehrz80}).
Taking $\theta_{bb}$ = 0.2 milliarcsec estimated above and a value of 600 km s$^{-1}$ observed for the FWHM of the emission line profile of Pa$\beta$ line in 
the spectrum taken on Nov. 3.59 UT 2009 for $V_{ej}$, we get
d = 24 kpc. This value for d is about a factor of 4 larger than the distance derived earlier in section 3.1 and indicates that the psudo-photosphere behaves
like a grey body with reduced  emissivity (see \cite{Das08}, \cite{Raj11}).

\begin{figure}
\begin{center}
\includegraphics[width=3.3in,height=3.5in,clip]{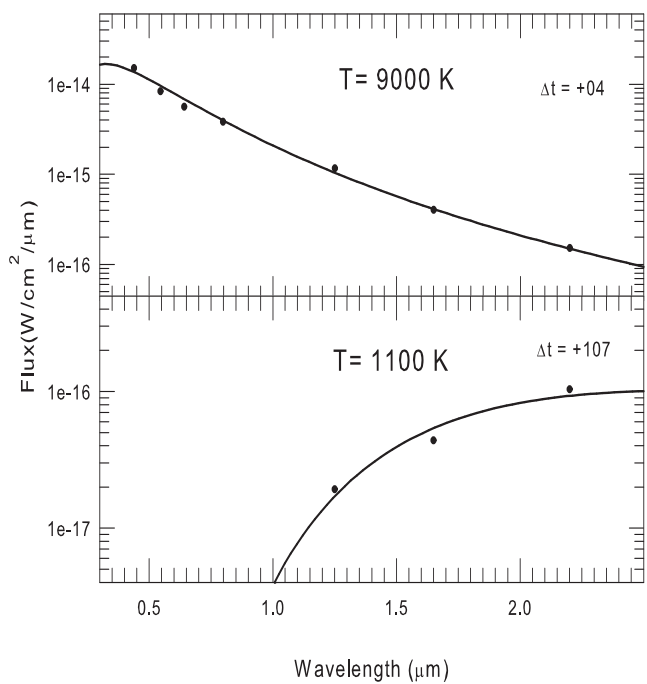}
\caption[The $V$ and $JHK$ band light curves of KT Eri]{The top panel shows SED for the fireball phase data of Nov. 2 UT 2009 with blackbody temperature 
fit of 9000 K. The bottom panel shows a blackbody fit to the data taken on Feb. 14 UT 2010, with a temperature of about 1100 K.}
\label{ch5_2}
\end{center}
\end{figure}

\subsection{Dust formation and ejecta mass estimate}

The detection of the CO first overtone emission in the spectrum taken on Nov. 5.64 UT 2009 is consistent with the detection of the dust in the nova ejecta 
reported by
\cite{Russell10} on Feb. 10 UT 2010. Although a decline in brightness at 
optical wavelengths is expected at the 
time of dust formation, the lack of observations from $\sim$ 30 till $\sim$ 110 days after the discovery 
due to the solar conjuction is the likely reason for absense of a sharp fall in the $V$ band light curve
shown in Fig. 1. The dust 
formation most likely took place during this time period. 
The thermal emission from the dust contributes to the near-IR bands and one expects a brightening at these 
wavelengths.Although our NIR photometric 
observations presented in Fig. 2 do not cover the dust formation phase, data is available 
in the Stony Brook database (SMARTS Spectral Atlas of Southern Novae) which
clearly shows a brightening in the $K$  band from Feb. 14 UT 2010 
($www.astro.sunysb.edu/fwalter/SMARTS/NovaAtlas/$,\cite{Walter12}). From this database, 
using $JHK$ magnitudes obtained on Feb. 14 UT 2010 (11.36 mag in J, 9.05 mag in H and 6.78 mag in K), we estimate a value of 1100 $\pm$ 200 K for the dust shell temperature from 
the SED plot shown in the lower panel of
Fig. 8. It is possible that the thermal emission from the dust, which is seen to be increasing upto the $K$ band, peaks at even longer wavelengths.

An approximate estimate of the mass of the dust shell can be made from the thermal component of the SED of Feb. 14 UT 2010 shown in Fig. 8 (lower panel).
Using the \cite{Woodward93} relation, we have calculated the mass of the dust shell, viz,

\begin{equation}\nonumber
M_{dust} = 1.1 \times 10{^{\rm 6}} (\lambda F_{\lambda})_{max} d^2 / T_{dust}^6
\end{equation}

The mass of the dust shell $M_{dust}$ is in units of $M_\odot $, $(\lambda F_{\lambda})_{max}$ is in W cm$^{-2}$,
the black-body temperature of the dust shell T$_{dust}$ is in units of 10${^{\rm 3}}$ K, and the distance to the nova d is in kpc in above relation.

We obtain $M_{dust}$ = 7.35 $\times$ 10${^{\rm -9}}$ $M_\odot$ for Feb. 14 UT 2010 taking the observed parameters of
$(\lambda F_{\lambda})_{max}$ = 2.98 $\times$ 10${^{\rm -16}}$ W cm$^{-2}$, $T_{dust}$= 1.1 $\times$ 10${^{\rm 3}}$
 K and d = 6.3 kpc. Taking a  canonical  value of 200 for
 the gas-to-dust ratio, we get 1.5 $\times$ 10${^{\rm -7}}$  M$_\odot $ for the gaseous component of the ejecta. This value is smaller than the typically
 observed value of 10${^{\rm -6}}$ to 10${^{\rm -4}}$ M$_\odot$.

 The reasons for the lower mass estimate could be several. It is possible that dust condensation has occurred only in certain regions and not over the entire extent
 of the ejecta. Further, the black body temperatures may
not represent the actual dust temperatures in an accurate manner for two reasons. First, the emissivity of the dust grains depends on their composition and 
size distribution and the frequency dependence of the emissivity can deviate from that of a
blackbody (\cite{Kruegel03}). Second, the data used for fitting the SED covers only the 1 to 2.5 $\mu$m region resulting in overestimation of the dust 
temperature as emission at longer wavelengths is not considered. The dust mass is very sensitive to $T_{dust}$ and its correct estimate should result in enhanced 
dust mass. The lower dust temperature of 880 $\pm$ 50 K derived by \cite{Russell10} from the 3 to 14 $\mu$m spectroscopic observations indicate that there is 
significant contribution at longer wavelengths. The observations from the Wide field Infrared Survey Explorer (WISE) also support emission from the dust at 
longer wavelengths. WISE detects the source in all the 3.4 (W1), 4.6 (W2), 12 (W3) and 22 $\mu$m (W4) bands; the  emission at the longer W3, W4 bands is very 
pronounced.

A few notes on the WISE images in Fig. 9 are necessary. The W1 and W2 images were taken at two epochs separated by nearly six months apart namely on 
Mar. 27 and Sep. 28 UT 2010. These double-epoch images appear to have been combined for the resultant W1, W2 images available at the WISE portal and which are presented 
here in the
figure. For reasons that are not clear, in the process of combining the images certain artifacts have been created in the W2 image and the image of V5584 Sgr were
smeared out and also partially obliterated. This is the likely reason why none of the W1, W2, W3 or W4 magnitudes are reported for the source. 
On the other hand the W3 and W4 images were taken on two very nearby epochs namely on Mar. 27 and Apr. 06 UT 2010 which give 
a good qualitative idea that pronounced dust emission is present in these wavebands.

\begin{figure*}
\begin{center}
\includegraphics[width=5in,height=3.5in,clip]{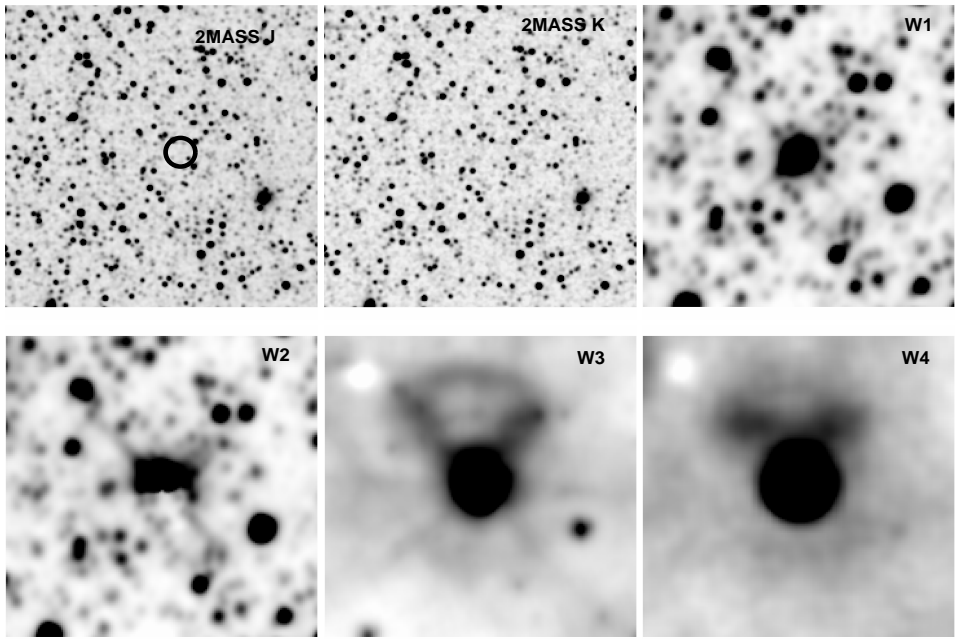}
\caption[Wise]{A mosaic of the same 4x4 arc minute square  field around V5584 Sgr. Shown are the
pre-outburst 2MASS J and K band images with the nova progenitor barely visible; its position is circled in the J band image. The WISE images detect the source in 
all the 3.4 (W1), 4.6 (W2), 12 (W3) and 22 $\mu$m (W4) bands; the  emission at the longer W3, W4 bands is very pronounced. The WISE images were taken in 
March-April 2010 (more details in section 3.5), nearly 5 months since discovery of Oct. 26.439 UT 2009. Although the nova had faded below 15 magnitudes 
in the $V$ band by this time (see Fig. 1), it remained strikingly bright in the near and mid IR due to emission from newly formed dust in the ejecta.}
\label{ch5_2}
\end{center}
\end{figure*}

\begin{table}
\caption[]{A list of all  Galactic Novae which have shown first-overtone CO emission.}
\begin{tabular}{lcl}
\hline\\
Nova & Detection epoch & Reference \\
&  after outburst  (days)& \\
\hline \\

NQ Vul               &         19    & \cite{Ferland79}\\
V842 Cen          &         25    &  \cite{Wichmann91}\\
V705 Cas          &           6    &  \cite{Evans96}  \\
V2274 Cyg        &         17    &  \cite{Rudy03}          \\
V2615 Oph        &          9     & \cite{Das09}\\
V5584 Sgr         &        12     & This work\\
V496 Sct           &        19     & \cite{Rudy09} $\&$ \cite {Raj12}\\
V2676 Oph       &        37     &  \cite{Rudy12a}   \\
V1724 Aql       &          7      &  \cite{Rudy12b} \\

\hline
\end{tabular}
\label{table4}
\end{table}

\section{Summary}

We have presented near-infrared spectroscopy and photometry of nova V5584 Sgr which erupted on Oct. 26.439 UT 2009.
From the optical light curve, V5584 Sgr is seen to be a moderately fast nova
with $t_2$ = 25 days and a light curve classification of D(46) following the classification scheme of \cite{Strope10}. 
The distance to the nova and its height below the Galactic plane are estimated to be 6.3 $\pm$ 0.5 kpc and 341 $\pm$ 30 pc,
respectively.
The outburst bolometric luminosity of V5584 Sgr as derived from its estimated M$_V$ is
L$_O$ $\sim$ 1.1 $\times$ 10$^5$ L$_\odot$. 
The infrared spectra indicates that the nova is of the Fe II type. The first overtone CO emission is the notable feature of the near-infrared spectrum
in the early decline phase.
The CO emission is modeled to make estimates of the CO mass and temperature. We discuss dust formation in the nova and make estimates of the mass of the dust 
and the gas  in the ejecta.

\section{Acknowledgements}

The research work  at the Physical Research Laboratory is funded by the Department of Space, Government of India.
 We are grateful for the availability of AAVSO (American Association of Variable Star Observers) optical photometric data, WISE near and mid-infrared data and 
 near-IR $JHK$ magnitudes from the STONY BROOK/SMARTS data collection. The authors are thankful to the anonymous refree for assiduous comments that improved 
 the manuscript.

\bibliographystyle{raa}
 \bibliography{bibtex1}

\begin{thebibliography}{33}
\providecommand{\natexlab}[1]{#1}
\providecommand{\selectlanguage}[1]{\relax}

\bibitem[{{Banerjee} \& {Ashok}(2012)}]{Banerjee12}
{Banerjee}, D.~P.~K., \& {Ashok}, N.~M. 2012, Bulletin of the Astronomical
  Society of India, 40, 243

\bibitem[{{Corelli}(2009)}]{Corelli09}
{Corelli}, P. 2009, Central Bureau Electronic Telegrams, 1994, 1

\bibitem[{{Das} et~al.(2009){Das}, {Banerjee}, \& {Ashok}}]{Das09}
{Das}, R.~K., {Banerjee}, D.~P.~K., \& {Ashok}, N.~M. 2009, \mnras, 398, 375

\bibitem[{{Das} et~al.(2008){Das}, {Banerjee}, {Ashok}, \& {Chesneau}}]{Das08}
{Das}, R.~K., {Banerjee}, D.~P.~K., {Ashok}, N.~M., \& {Chesneau}, O. 2008,
  \mnras, 391, 1874

\bibitem[{{della Valle} \& {Livio}(1995)}]{Della95}
{della Valle}, M., \& {Livio}, M. 1995, \apj, 452, 704

\bibitem[{{Evans} et~al.(1996){Evans}, {Geballe}, {Rawlings}, \&
  {Scott}}]{Evans96}
{Evans}, A., {Geballe}, T.~R., {Rawlings}, J.~M.~C., \& {Scott}, A.~D. 1996,
  \mnras, 282, 1049

\bibitem[{{Ferland} et~al.(1979){Ferland}, {Lambert}, {Netzer}, {Hall}, \&
  {Ridgway}}]{Ferland79}
{Ferland}, G.~J., {Lambert}, D.~L., {Netzer}, H., {Hall}, D.~N.~B., \&
  {Ridgway}, S.~T. 1979, \apj, 227, 489

\bibitem[{{Gehrz}(2008)}]{Gehrz08}
{Gehrz}, R.~D. 2008, {in Bode M.F., Evans A., eds, Classical Novae, 2nd Edn.
  Cambridge Univ. Press, Cambridge, p167}

\bibitem[{{Gehrz} et~al.(1980){Gehrz}, {Grasdalen}, {Hackwell}, \&
  {Ney}}]{Gehrz80}
{Gehrz}, R.~D., {Grasdalen}, G.~L., {Hackwell}, J.~A., \& {Ney}, E.~P. 1980,
  \apj, 237, 855

\bibitem[{{Kinugasa} et~al.(2009){Kinugasa}, {Honda}, {Hashimoto}, {Taguchi},
  \& {Takahashi}}]{Kinugasa09}
{Kinugasa}, K., {Honda}, S., {Hashimoto}, O., {Taguchi}, H., \& {Takahashi}, H.
  2009, Central Bureau Electronic Telegrams, 1995, 1

\bibitem[{{Kruegel}(2003)}]{Kruegel03}
{Kruegel}, E. 2003, {The Physics of Intersteller Dust, The Institute of Physics
  Ser. in Astron. \& Astrophys., Bristol}

\bibitem[{{Maehara}(2009)}]{Maehara09}
{Maehara}, H. 2009, Central Bureau Electronic Telegrams, 1995, 2

\bibitem[{{Munari} et~al.(2009){Munari}, {Saguner}, {Siviero}
  et~al.}]{Munari09}
{Munari}, U., {Saguner}, T., {Siviero}, A., et~al. 2009, Central Bureau
  Electronic Telegrams, 1999, 1

\bibitem[{{Ney} \& {Hatfield}(1978)}]{Ney78}
{Ney}, E.~P., \& {Hatfield}, B.~F. 1978, \apjl, 219, L111

\bibitem[{{Nishiyama} et~al.(2009){Nishiyama}, {Kabashima}, \&
  {Corelli}}]{Nishiyam09}
{Nishiyama}, K., {Kabashima}, F., \& {Corelli}, P. 2009, Central Bureau
  Electronic Telegrams, 1994, 1

\bibitem[{{Poggiani}(2011)}]{Poggiani11}
{Poggiani}, R. 2011, \apss, 333, 115

\bibitem[{{Pontefract} \& {Rawlings}(2004)}]{Pontefract04}
{Pontefract}, M., \& {Rawlings}, J.~M.~C. 2004, \mnras, 347, 1294

\bibitem[{{Raj} et~al.(2011){Raj}, {Ashok}, \& {Banerjee}}]{Raj11}
{Raj}, A., {Ashok}, N.~M., \& {Banerjee}, D.~P.~K. 2011, \mnras, 415, 3455

\bibitem[{{Raj} et~al.(2009){Raj}, {Ashok}, {Banerjee}, \& {Hornoch}}]{Raj09}
{Raj}, A., {Ashok}, N.~M., {Banerjee}, D.~P.~K., \& {Hornoch}, K. 2009, Central
  Bureau Electronic Telegrams, 2002, 1

\bibitem[{{Raj} et~al.(2012){Raj}, {Ashok}, {Banerjee} et~al.}]{Raj12}
{Raj}, A., {Ashok}, N.~M., {Banerjee}, D.~P.~K., et~al. 2012, \mnras, 425, 2576

\bibitem[{{Rawlings}(1988)}]{Rawlings88}
{Rawlings}, J.~M.~C. 1988, \mnras, 232, 507

\bibitem[{{Rudy} et~al.(2003){Rudy}, {Dimpfl}, {Lynch} et~al.}]{Rudy03}
{Rudy}, R.~J., {Dimpfl}, W.~L., {Lynch}, D.~K., et~al. 2003, \apj, 596, 1229

\bibitem[{{Rudy} et~al.(2012{\natexlab{a}}){Rudy}, {Laag}, {Crawford}
  et~al.}]{Rudy12b}
{Rudy}, R.~J., {Laag}, E.~A., {Crawford}, K.~B., et~al. 2012{\natexlab{a}},
  Central Bureau Electronic Telegrams, 3287, 1

\bibitem[{{Rudy} et~al.(2009){Rudy}, {Prater}, {Puetter}, {Perry}, \&
  {Baker}}]{Rudy09}
{Rudy}, R.~J., {Prater}, T.~R., {Puetter}, R.~C., {Perry}, R.~B., \& {Baker},
  K. 2009, \iaucirc, 9099, 1

\bibitem[{{Rudy} et~al.(2012{\natexlab{b}}){Rudy}, {Russell}, {Sitko}
  et~al.}]{Rudy12a}
{Rudy}, R.~J., {Russell}, R.~W., {Sitko}, M.~L., et~al. 2012{\natexlab{b}},
  Central Bureau Electronic Telegrams, 3103, 1

\bibitem[{{Russell} et~al.(2010){Russell}, {Laag}, {Rudy}, {Skinner}, \&
  {Gregory}}]{Russell10}
{Russell}, R.~W., {Laag}, E.~A., {Rudy}, R.~J., {Skinner}, M.~A., \& {Gregory},
  S.~A. 2010, \iaucirc, 9118, 2

\bibitem[{{Schlafly} \& {Finkbeiner}(2011)}]{Schlafly11}
{Schlafly}, E.~F., \& {Finkbeiner}, D.~P. 2011, \apj, 737, 103

\bibitem[{{Strope} et~al.(2010){Strope}, {Schaefer}, \& {Henden}}]{Strope10}
{Strope}, R.~J., {Schaefer}, B.~E., \& {Henden}, A.~A. 2010, \aj, 140, 34

\bibitem[{{Walter} et~al.(2012){Walter}, {Battisti}, {Towers}, {Bond}, \&
  {Stringfellow}}]{Walter12}
{Walter}, F.~M., {Battisti}, A., {Towers}, S.~E., {Bond}, H.~E., \&
  {Stringfellow}, G.~S. 2012, \pasp, 124, 1057

\bibitem[{{Warner}(1995)}]{Warner95}
{Warner}, B. 1995, Cataclysmic Variable Stars, Cambridge Astrophys. Ser.
  Cambridge Univ. Press

\bibitem[{{Warner}(2008)}]{Warner08}
{Warner}, B. 2008, {in Bode M.F., Evans A., eds, Classical Novae, 2nd Edn.
  Cambridge Univ. Press, Cambridge, p21}

\bibitem[{{Wichmann} et~al.(1991){Wichmann}, {Krautter}, {Kawara}, \&
  {Williams}}]{Wichmann91}
{Wichmann}, R., {Krautter}, J., {Kawara}, K., \& {Williams}, R.~E. 1991, in The
  Infrared Spectral Region of Stars, edited by C.~{Jaschek} \& Y.~{Andrillat},
  353

\bibitem[{{Woodward} et~al.(1993){Woodward}, {Lawrence}, {Gehrz}
  et~al.}]{Woodward93}
{Woodward}, C.~E., {Lawrence}, G.~F., {Gehrz}, R.~D., et~al. 1993, \apjl, 408,
  L37

\end{thebibliography}

\end{document}